# TRANSPORT AND MAGNETIC PROPERTIES

# OF LT ANNEALED Ga$_{1-x}$Mn$_x$As

I. Kuryliszyn [a], T. Wojtowicz [b], X. Liu [b], J.K. Furdyna [b], W. Dobrowolski [a],

[a] Institute of Physics, Polish Academy of Sciences, Warsaw, Poland

[b] Department of Physics, University of Notre Dame, Notre Dame, USA

J.-M. Broto, M. Goiran, O. Portugall, H. Rakoto, B. Raquet

Laboratoire National des Champs Magnétiques Pulsés Toulouse, France

We present the results of low temperature (LT) annealing studies of Ga$_{1-x}$Mn$_x$As epilayers grown by low temperature molecular beam epitaxy in a wide range of Mn concentrations ($0.01 < x < 0.084$). Transport measurements in low and high magnetic fields as well as SQUID measurements were performed on a wide range of samples, serving to establish optimal conditions of annealing. Optimal annealing procedure succeeded in the Curie temperatures higher than 110K. The highest value of Curie temperature estimated from the maximum in the temperature dependence of zero-field resistivity ($T_\rho$) was 127K. It is generally observed that annealing leads to large changes in the magnetic and transport properties of GaMnAs in the very narrow range of annealing temperature close to the growth temperature.

PACS numbers: 75.50.Pp, 75.50.Dd, 81.40.Rs

## 1. Introduction

Ferromagnetic semiconductors have recently received much interest, since they hold out prospects for using electron spins in electronic devices for the processing, transferring, and storing information [1,2]. Particularly $Ga_{1-x}Mn_xAs$ has become the focus of current interest because of its high Curie temperature ($T_C \sim$ 110K) and possible spin-electronics applications [2,3,4]. It is widely accepted that ferromagnetic behavior of GaMnAs is connected with its p-type nature [4,5,6]. The Mn ions incorporated into the III-V semiconductor matrix play the dual role of acceptor sites and magnetic ions. The ferromagnetic ordering of the Mn magnetic moments is mediated by free holes.

More recently, it was reported that heat treatment (annealing) of the grown by low-temperature molecular beam epitaxy (MBE) GaMnAs improves the Curie temperature, magnetization M(T) and $T_C$ of the annealed samples depending on both the annealing temperature and the duration of the annealing process [7,8].

We present the results of magnetic measurements and magnetotransport investigations performed in the range of low and high magnetic fields of as-grown and annealed $Ga_{1-x}Mn_xAs$ samples. We have grown and systematically studied the $Ga_{1-x}Mn_xAs$ samples over a wide range of Mn concentration: 0.01<x<0.084.

## 2. Growth method and experimental

The layers of $Ga_{1-x}Mn_xAs$ were grown using low temperature (LT) MBE. Semi-insulating epiready (100) GaAs wafers were used as substrates. Typically, a buffer of GaAs was first grown at high temperature ($600^0$C). The substrate was then cooled to a temperatures in the range $T_S \sim 250^0$C $- 275^0$C, and a layer of low temperature (LT) GaAs was grown to a thickness in the range between 2nm -100nm. Finally $Ga_{1-x}Mn_xAs$ layer in the same range of substrate temperatures to a thickness of the 105 nm $-$ 302 nm was grown. The growth was monitored *in situ* by reflection high energy electron diffraction (RHEED). The lattice constant of GaMnAs samples was measured by x-ray diffraction (XRD). The Mn concentration x was determined using two different methods. First, the Mn content was obtained from XRD measurements by assuming that the GaMnAs layer is fully strained by the GaAs substrate. And second, during the growth the x values were estimated from the change in the growth rate monitored by RHEED oscillations. The results determined from these two methods were in good agreement.

Next, the as-grown wafers were cut into a number of specimens for systematic annealing experiments. The samples were annealed at low temperatures in the range between $260^0$C and $350^0$C. The optimal time of annealing was between 1h and 1.5 h, in agreement with the results of Potashnik et. al. [8]. All the samples were annealed under the same fixed flow of $N_2$ gas of 1.5 SCFH (standard cubic feet per hour). All post-growth and post-annealing procedures were carried out in the same manner. After annealing, the samples were cooled down by a rapid quench under the flow of nitrogen gas. For all the samples the electrical contacts for the transport measurements were also prepared always in the same way. We used Hall bar samples with typical dimensions of 2mm $\times$ 6mm. The electrical contacts were made using indium solder and gold wires. The optimal annealing temperature turned out to be near the growth temperature (i.e., around $280^0$C) in the case of all $Ga_{1-x}Mn_xAs$ samples with Mn concentration exceeding x $\approx$ 0.08.

The annealed samples were characterized by means of resistivity measurements at zero magnetic field in the temperature range between 10K and room temperature using a helium flow cryostat. The temperature dependence of the zero-field resistivity shows a hump structure (see Fig 1.). Such feature is known to appear at the temperature slightly above the value of $T_C$ obtained from magnetic measurements. The magnetization measurements M(T) revealed good agreement between transport and magnetic results. The magnetization was measured by a SQUID magnetometer in small magnetic fields (i.e., 10 Gauss) after the sample has been magnetized at higher magnetic field (1000 gauss) parallel to the sample surface. For the epilayer annealed at the optimal temperature of $289^0$C, the value corresponding to the zero-field resistivity hump is equal to 127K, and the SQUID measurements on the same sample yielded $T_C$=119K.

Fig. 2 presents the Curie temperature estimated from the maximum in the temperature dependence of the zero-field resistivity ($T_\rho$) as a function of the annealing temperature. We observe large changes in the Curie temperature in a very narrow range of annealing temperatures. The great enhancement of the Curie temperature in samples with a high Mn content, $x \geq 0.05$, is clearly visible. For the sample with x=0.083, $T_\rho$ shifts from 88K for the as-grown sample to the 127K after annealing at the optimal temperature of $289^0C$ and drops to $T_\rho$=30K after heat treatment at $350^0C$. Fig. 1 shows typical temperature dependences of the zero-field resistivity for the $Ga_{1-x}Mn_xAs$ epilayer with x=0.083 for various annealing procedures. The resistivity of the as-grown sample and for samples annealed at relatively low temperatures ($260^0C$, $280^0C$, $300^0C$ and $310^0C$) show typical metallic behavior. For samples annealed at $350^0C$,  an insulating behavior of the resistivity is observed.

The increase of the Curie temperature is always accompanied by the enhancement of the conductivity. We observe also an increase of the saturation magnetization for the samples annealed at the optimal conditions as compared to the as-grown specimens.

In the case of low Mn concentration, $x < 0.05$, the influence of the annealing procedure on both the Curie temperature and conductivity is weak.

One of the annealed and characterized samples with high content of Mn ($x \sim 0.08$) was additionally investigated by means of channeling Rutherford backscattering (c-RBS) and channeling particle induced X-ray emission (c-PIXE) experiments [9]. The results of channeling measurements indicate that LT annealing introduces a rearrangement of Mn sites in the $Ga_{1-x}Mn_xAs$ lattice. In particular, the large increase of $T_C$, accompanied by the increase of saturation magnetization and free carrier concentration (measured by electrochemical capacitance-voltage method) can be attributed to the relocation of Mn atoms from interstitial ($Mn_I$) to substitutional sites ($Mn_{Ga}$) or to random positions (which Mn can occupy if it precipitates in the form of other phases, e.g., as MnAs). The removal of a significant fraction of interstitial Mn ($Mn_I$) provides a clear explanation of the three effects observed after annealing, i.e., the increase of the hole concentration, the increase of the Curie temperature, and the increase in the saturation magnetization observed at low temperatures. First, interstitial Mn atoms (having two valance electrons) act as double donors, thus compensating the substitutional Mn ($Mn_{Ga}$) acceptors. Removal of $Mn_I$ therefore increases the number of electrically-active $Mn_{Ga}$, and thus also the hole concentration. As is well known, the increase in the hole concentration will automatically result in an increase of $T_C$. Finally, one should realize that the interstitial $Mn_I$ donor is both positively charge and relatively mobile. It is therefore expected to drift toward the negatively charged $Mn_{Ga}$ acceptor centers, thus forming $Mn_{Ga}$-$Mn_I$ pairs. Because removal of $Mn_I$ by annealing is accompanied by an increase of saturation magnetization (as evidenced by our magnetization studies), we infer that the $Mn_{Ga}$-$Mn_I$ pairs are coupled antiferromagnetically; i.e., the magnetic moment of $Mn_I$ "neutralizes" the contribution of $Mn_{Ga}$ to the magnetization [10]. Removal of $Mn_I$ from such a pair should thus automatically render the substitutional $Mn^{++}$ magnetically-active, increasing the saturation magnetization, as is indeed observed experimentally.

We have studied the hysteresis loops using SQUID magnetometer of the as-grown and annealed samples of GaMnAs. In addition, the Hall resistivity and conductivity in the range of high-pulsed magnetic fields up to 30T were also measured.

Ferromagnetic GaMnAs epilayers are characterized by the presence of the anomalous Hall effect (AHE). Determination of the free carrier concentration is therefore complicated by the dominance of the anomalous Hall effect term. In principle, the transport investigations should be performed at low temperatures and at magnetic fields sufficiently high so that the magnetization saturates. We have noted that the magnetotransport measurements (Hall resistivity and conductivity) that we performed up to 30T and at low temperatures do not give a unique value for the hole concentration of the investigated epilayers, indicating that 30T is insufficient to saturate the AHE contribution. Much higher magnetic fields are therefore required to determine the free carrier concentration of the investigated GaMnAs epilayers.

However we observe a very pronounced decrease of the magnetoresistivity for the samples annealed at the optimal temperature (see Fig. 3).

The measurements of M(B) showed that the annealing process also affects the hysteresis loops specifically, we observed that the coercive field $H_C$ decreases when the samples are annealed at the optimal temperatures around $280^0$C. This effect is shown in Fig. 4. Simultaneously, the saturation magnetization $M_S$ increases after heat treatment at the optimal conditions, indicating that annealing increases the concentration of magnetically-active Mn ions.

### 3. Conclusions

In summary, we have systematically studied both the magnetic and transport properties of the annealed $Ga_{1-x}Mn_xAs$ epilayers. We observe a large enhancement of ferromagnetism for the samples annealed at an optimal temperature, typically about $280^0$C. The increase of the Curie temperature is accompanied by an increase of conductivity and saturation magnetization, also the increase of the hole concentration [9]. Such large changes of the magnetic and transport properties induced by low temperature annealing can be attributed to the relocations of Mn atoms in the GaMnAs lattice [9].

Annealing at optimal temperatures also leads to a decrease of the coercive field $H_c$ and of magnetoresistivity.

Presently, only very speculative explanation of these experimental results is possible. It is very likely that the observed features of hysteresis loop such as the coercive field, the shape of the loop, and the value of the remanent magnetization and of the saturation magnetization are associated with the magnetic domain structure. The size and the shape of the magnetic domains reflect the magnitude and the anisotropy of the microscopic exchange interaction.

Recently, the unconventional random domain structure in GaMnAs films with in-plane magnetization have been reported [11]. Moreover, Barabash and Stroud [12] showed that both the shape and the positions of the peaks in the magnetoresistivity depend on the domain microgeometry and on the squareness of the hysteresis loop. We suggest that change of the domain structure can lead to the experimental results reported in this paper. However the rigorous interpretation of the experimental results is difficult, because of the absence of understanding of the fundamentals of ferromagnetism in GaMnAs layers.

**Acknowledgments**

This work was partially supported within European Community program ICA1-CT-2000-70018 (Center of Excellence CELDIS) and DARPA SpinS Program. T.W. was supported by the Fulbright Foundation it the form of Senior Fulbright Fellowship.

[1] G. A. Prinz, Science **282**, 1660 (1998)

[2] H. Ohno, Science **281**, 951 (1998)

[3] T. Dietl, H. Ohno, F. Matsukura, J. Cibert and D. Ferrand, Science **287** (2000)

[4] "Ferromagnetic III-V Semiconductors", F. Matsukura, H. Ohno, T. Dietl in: Handbook on
     Magnetic Materials, Elsevier 2002 in press.

[5] F. Matsukura, H. Ohno, A.Shen and Y. Sugawara, Phys. Rev. B **57**, R2037 (1998)

[6] H. Ohno, J. Magn. Magn. Mater. **200**, 110 (1999)

[7] T. Hayashi, Y. Hashimoto, S. Katsumoto, and Y. Iye, Appl. Phys. Lett. **78**, 1691 (2001)

[8] S. J. Potashnik, K. C. Ku, S. H. Chun, J. J. Berry. N. Samarth, and P. Schiffer, Appl. Phys.
     Lett. **79**, 1495 (2001)

[9] K. M. Yu and W. Walukiewicz, T. Wojtowicz, I. Kuryliszyn, X. Liu, Y. Sasaki, and J. K.
     Furdyna,  Phys. Rev. **B 65**, 201303(R) (2002)

[10] J. Blinowski, P. Kacman, K.M. Yu, W. Walukiewicz, T. Wojtowicz and J.K. Furdyna,
      to be published

[11] T. Fukumura, T. Shono, K. Inaba, T. Hasegawa, H. Koinuma, F. Matsukura, H. Ohno,
      Physica E **10**, 135, (2001)

[12] S. V. Barabash  and D. Stroud, Appl. Phys. Lett **79**, 979, (2001)

**Figure Caption**

**Fig. 1.** Temperature dependence of the zero-field resistivity of GaMnAs samples with high Mn concentration (x=0.083) annealed at various temperatures.

**Fig. 2.** The temperature $T_\rho$ versus temperatures of annealing for different Mn concentration: (8.3%■, 8.1%σ, 8.4%●, 6.1%□, 3.2%▨, 1.5%▨).

**Fig. 3.** Magnetization measured as a function of magnetic field for epilayers of GaMnAs with x=0.081, either as-grown or annealed at the optimal conditions. Note that the coercive field decreases for the sample annealed at an optimal temperature of $T_a=282^0$C.

**Fig. 4.** Magnetoresistivity (R-$R_0$/$R_0$, where $R_0$ is the value of the resistivity at B=0) for two samples with x=0.083,: as-grown, and annealed at $289^0$C. The distinct decrease of the magnetoresistivity for the annealed sample is clearly seen in the data.

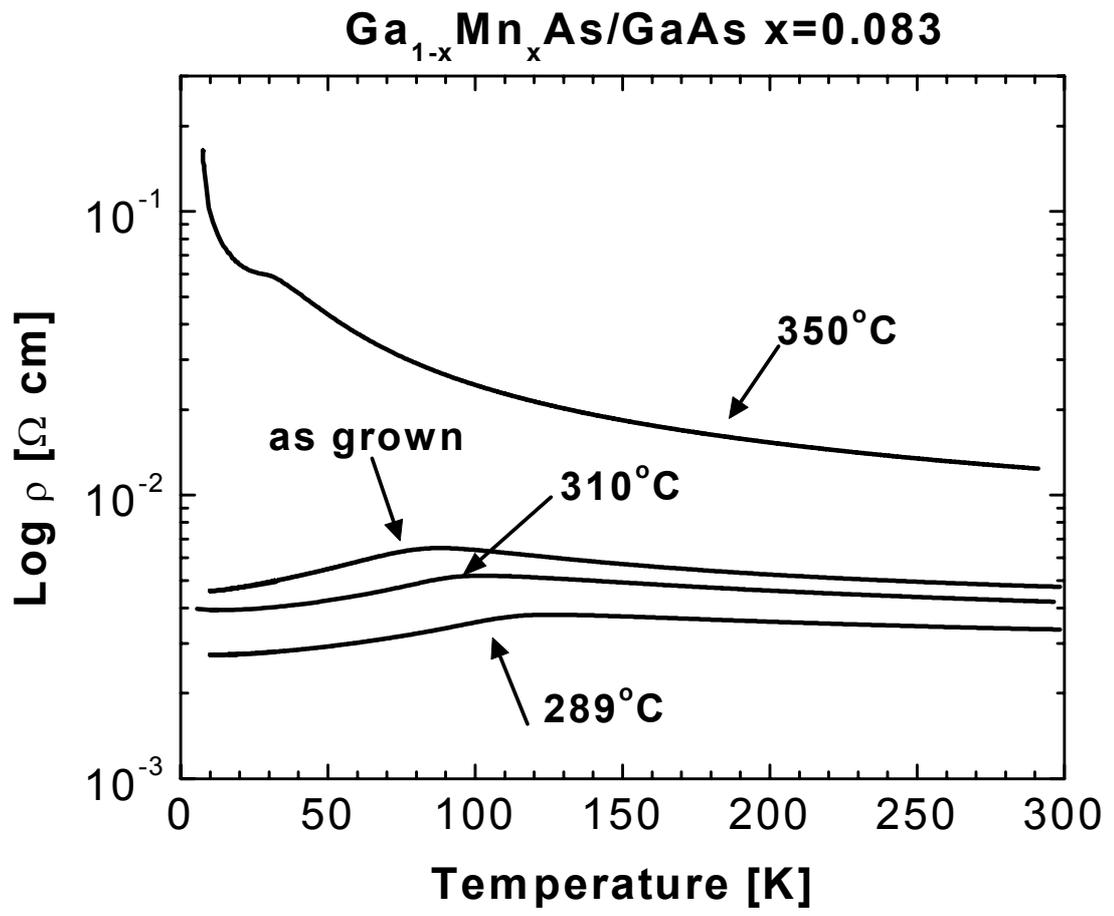

**Ga$_{1-x}$Mn$_x$As/GaAs x=0.083**

350°C

as grown

310°C

289°C

**Log ρ [Ω cm]**

**Temperature [K]**

Fig.1

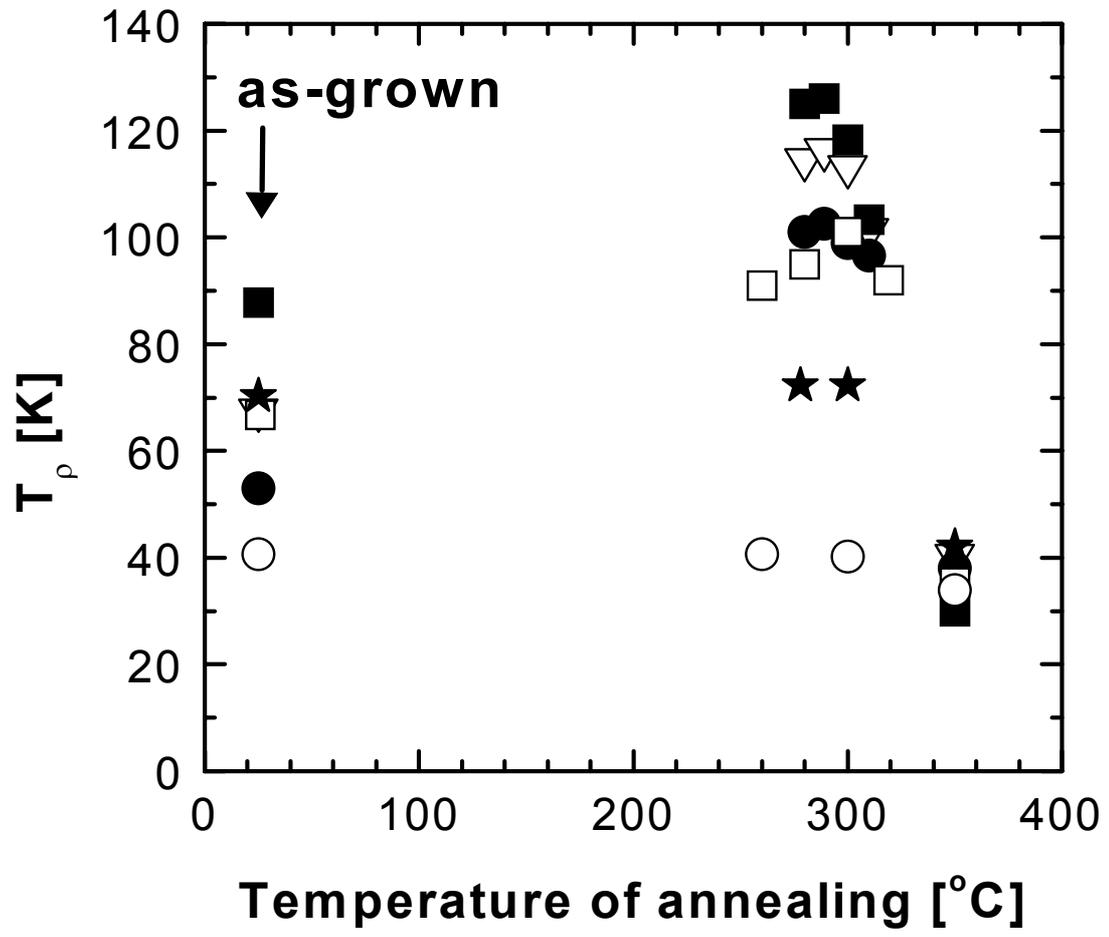

Fig.2

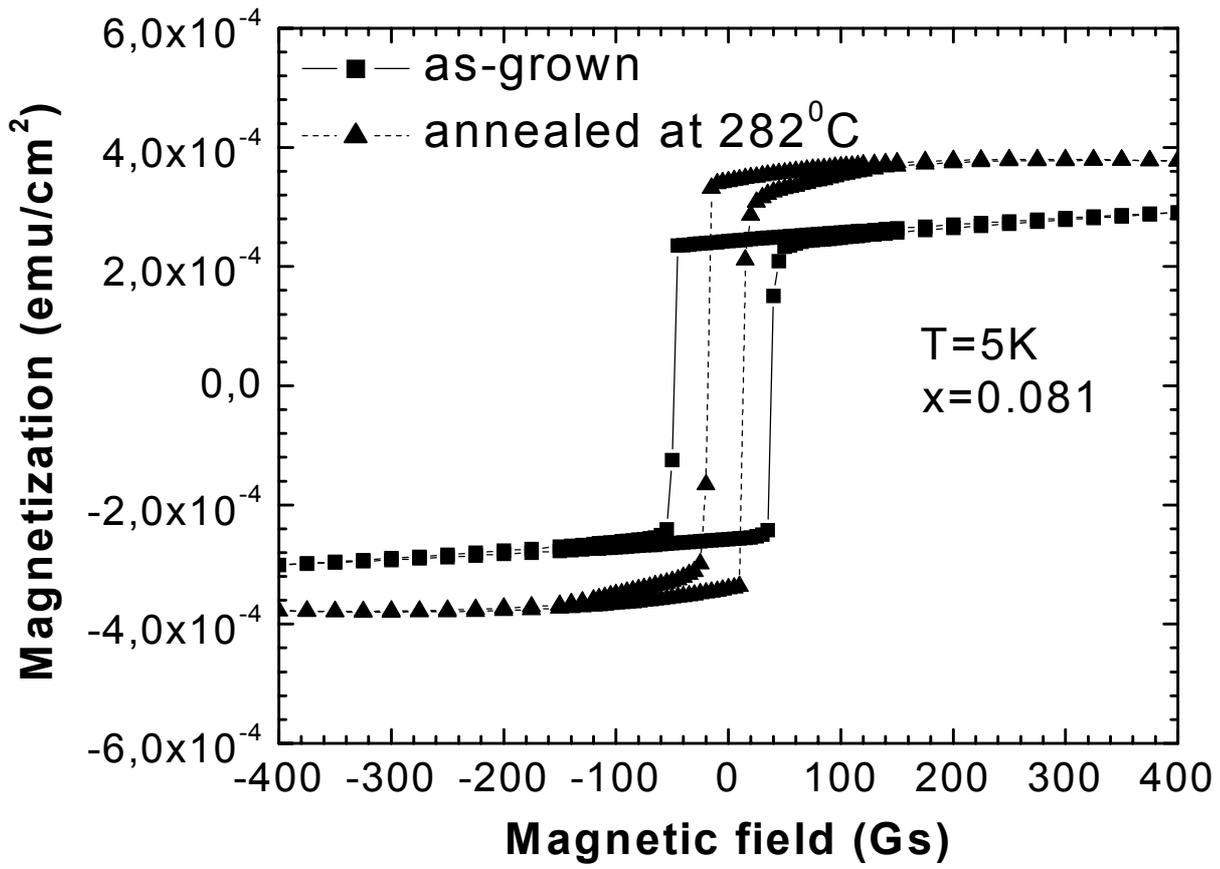

Fig.4

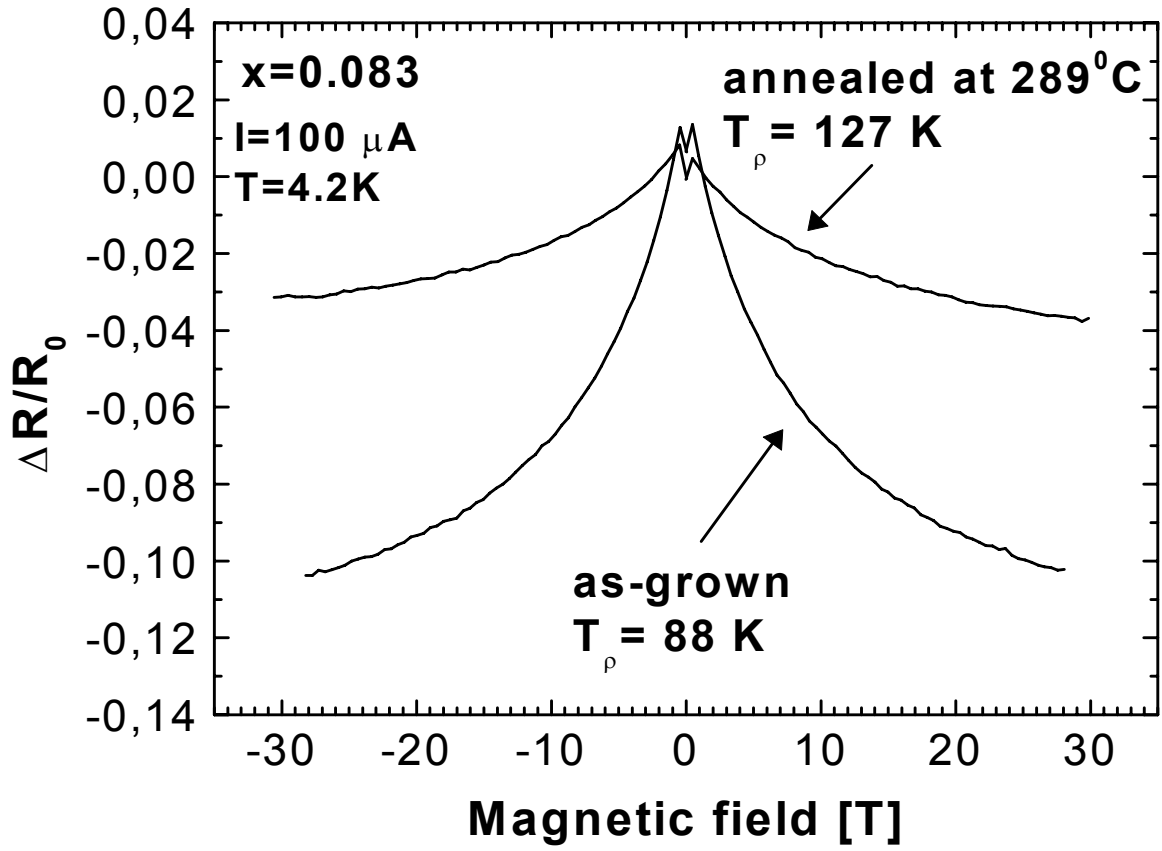

Fig.3